\documentclass[eqsecnum]{scrartcl}
\usepackage[T1]{fontenc}
\usepackage[latin1]{inputenc}
\usepackage{a4}
\usepackage{graphicx}
\usepackage{amssymb}

\makeatletter

\providecommand{\tabularnewline}{\\}

 \newcommand{\lyxaddress}[1]{
   \par {\raggedright #1 
   \vspace{1.4em}
   \noindent\par}
 }

\usepackage{graphicx}
\usepackage{graphicx}
\usepackage{graphicx}
\usepackage{bm}

\makeatother
\begin{document}

\title{Quantum dynamics with stochastic gauge simulations }

\author{P. D. Drummond and P. Deuar}

\maketitle

\lyxaddress{Australian Centre for Quantum-Atom Optics, The University of Queensland, St Lucia 4072, Queensland, Australia.}

\begin{abstract}
The general idea of a stochastic gauge representation is introduced and compared with more traditional phase-space expansions,
like the Wigner expansion. Stochastic gauges can be used to obtain an infinite class of positive-definite stochastic time-evolution
equations, equivalent to master equations, for many systems including quantum time-evolution. The method is illustrated with
a variety of simple examples ranging from astrophysical molecular hydrogen production, through to the topical problem of
Bose-Einstein condensation in an optical trap and the resulting quantum dynamics. 
\end{abstract}

\section{Introduction}

The original goal of pioneering physicists like Galileo, Newton and Einstein was to predict dynamical behavior in the universe
- from projectiles to planets and even photons. However, complex systems in nature are not soluble with analytic or direct
computational methods if the space of system descriptions is too large. Examples are master equations - widely used in many
disciplines - or quantum theory itself, where the Hilbert space for many-body systems is enormous. This is one of the central
problems of modern theoretical physics.

The problem of complexity inspired a claim by Feynman\cite{feynman:82} that reads;

\begin{itemize}
\item {}``Can a quantum system be probabilistically simulated by a classical universal computer? ...the answer is certainly, \textbf{No}!
'' (Richard P. Feynman, \emph{Simulating Physics with Computers,} 1982) 
\end{itemize}
In this paper, we will give a very general overview of new techniques for probabilistic simulation, called stochastic gauge
representations\cite{GaugeP}, which allow these `impossible' simulations. Like Wigner's original representation\cite{Wig-Wigner},
these are phase-space representations, but in phase-spaces of larger than classical dimension. The purpose of the paper is
to explain the abstract ideas behind stochastic gauges, which are essentially an equivalence class of curved-space path-integrals
in a complex phase-space. We compare the method with other phase-space approaches, as well as giving elementary examples.

Even in Feynman's day, the introduction of Quantum Monte Carlo (QMC) \cite{wilson:74} and related path integral techniques
indicated a possible way to solve problems of large Hilbert space dimension. While these methods are used to treat canonical
ensembles, they cannot treat real-time dynamics. Presumably, it was the failure of QMC methods in real time - due to the
highly oscillatory behaviour of the real-time Feynman path-integral - that caused Feynman to neglect these types of random
sampling methods.

In a more recent publication, the time-domain quantum simulation problem was restated by Ceperley\cite{Ceperley}, as follows:

\begin{itemize}
\item {}``There are serious problems in calculating the dynamics of quantum systems'' (David M. Ceperley, \emph{Microscopic simulations
in physic}s, 1999). 
\end{itemize}
The problems are due to the astronomical size of many-body Hilbert spaces. This makes it difficult to treat the quantum dynamics
of Bose-Einstein condensates - which typically have $10000$ or more particles, and $10^{10000}$ states in Hilbert space.
In this paper, we show that stochastic gauge methods are versatile enough to treat complex quantum systems including master
equations and canonical ensembles, as well as many-body quantum dynamics in real time.

\section{Phase-space representations}

Phase-space mappings, which map the discrete states of quantum theory into a classical-like phase-space, were originally
introduced by Wigner in the form of the famous Wigner representation\cite{Wig-Wigner}. In the stochastic version of these
methods, ensemble averages are mapped into trajectory averages - which can be numerically simulated. Phase-space representations
have developed in three distinct stages. In the first stage, the Wigner representation, Husimi\cite{Hus-Q} Q-representation
and Glauber-Sudarshan\cite{Gla-P,Sud-P} P-representation all use a classical phase-space of $2M$ real dimensions for quantum
systems corresponding to $M$ classical modes. These methods typically will not give locally positive propagators for nonlinear
quantum systems. Hence they cannot have a directly equivalent stochastic process.

In the second stage, higher dimensional representations were developed. These include the Glauber R-representation\cite{Gla-P}
(which is non-positive), the Poisson representation (for incoherent master equations only), and for quantum systems, the
positive P-representation\cite{+P}. The last two methods give locally positive propagators, by virtue of using a non-classical
phase-space. However, while they work very well for damped systems, in highly nonlinear simulations they typically develop
unstable trajectories\cite{SG-abs,SS-fail} leading to large sampling errors or even systematic errors in nonlinear quantum
simulations. These simulation problems are caused by the distributions having power-law tails in phase-space\cite{GGD-Validity}.
When this happens, the distributions are not sufficiently localized to allow integration by parts, which is a crucial step
in the derivation.

In the current stage, gauge representations are used, which add a further complex gauge amplitude \textbf{$\,\Omega$} to
the representation phase-space. These provide a way to overcome Feynman's complexity dilemma, and are the subject of the
present paper. For appropriate gauge choices these methods are conjectured to be exact in principle. They have no instabilities,
and we will show that numerical simulations give no systematic errors, even in cases where previous methods would have boundary
term errors. Numerical implementations are still necessarily approximate due to the usual limitations of finite computers,
but error levels can be estimated and reduced to any desired level compatible with the hardware and time available.

To understand the reason for these developments, it is useful to enumerate the requirements for a phase-space representation
that can be used to map quantum dynamics into a stochastic differential equation, which can be simulated numerically. They
are as follows:

\begin{enumerate}
\item \textbf{\textit{Non-singular:}} Essential \textit{\emph{for finite probabilities, although an initial delta function can
be tolerated.}}
\item \textbf{\textit{Positive distribution:}} Needed to get positive initial probabilities. 
\item \textbf{\emph{UV convergent:}} To control sampling error on lattices, vacuum fluctuations should not diverge at large momentum
cut-off. 
\item \textbf{\textit{2nd-order derivatives:}} The mapping must generate at most second-order derivatives, to obtain diffusive
phase-space behavior. 
\item \textbf{\textit{Positive-definite propagator:}} The short-time propagator must be positive-definite, for an equivalent stochastic
process to exist. 
\item \textbf{\textit{Stable:}} Trajectories in phase-space should be stable to prevent boundary term errors: further restrictions
on noise growth are also needed. 
\end{enumerate}
How do the known phase-space representations compare? This is shown in Table 1, for a system of much current interest, the
anharmonic oscillator --- See Sec.~\ref{anharm}.%
\begin{table}
\begin{center}\begin{tabular}{|c|c|c|c|c|c|c|}
\hline 
Phase-space&
 \textbf{Non-}&
 \textbf{Pos. }&
 \textbf{UV}&
 \textbf{2nd }&
 \textbf{Pos. }&
 \textbf{Stable?}\tabularnewline
 Repn.&
 \textbf{singular?}&
 \textbf{dist?}&
 \textbf{conv?}&
 \textbf{deriv?}&
 \textbf{def?}&
\tabularnewline
\hline
Wigner&
 Yes&
 No&
 No&
 No&
 No&
 -\tabularnewline
\hline
Q&
 Yes&
 Yes&
No &
 Yes&
 No&
 -\tabularnewline
\hline
P&
 No&
 No&
 Yes&
Yes &
 No&
 -\tabularnewline
\hline
R&
 Yes&
 No&
 Yes&
 Yes&
No &
 -\tabularnewline
\hline
Pos. P&
 Yes&
 Yes&
 Yes&
 Yes&
 Yes&
 Indefinite\tabularnewline
\hline
\textbf{Gauge}&
 \textbf{Yes}&
 \textbf{Yes}&
 \textbf{Yes}&
 \textbf{Yes}&
 \textbf{Yes}&
 \textbf{Yes  }\tabularnewline
\hline
\end{tabular}\end{center}

\caption{Table of representation properties required for stochastic quantum dynamics}
\end{table}

It can be easily seen that the earlier phase-space techniques using a classical phase-space had other problems as well as
not generating positive propagators. While the positive-P method removes almost all the problems of earlier techniques, it
still has a disadvantage in that it can generate moving singularities from unstable phase-space trajectories. This is known
to cause problems with large sampling errors, and boundary-terms\cite{SG-abs,SS-fail,GGD-Validity} causing systematic errors
in simulations of systems with extremely low damping.

The gauge representation method\cite{GaugeP}, which unifies and extends some other closely related approaches\cite{Paris1,Paris2,Plimak,ccp2k},
removes this last problem by stabilizing phase-space trajectories. It also points the way to the development of other phase-space
representations in future, since the fundamental idea relies on very general properties of completeness, analyticity, and
the existence of operator mappings into a second-order differential equation.

If we extend our table to other cases, we see that some phase space representations appear more suited to calculations other
than stochastic simulations. For example the symplectic tomography scheme of Mancini, Man'ko, and Tombesi\cite{Mancini},
which expresses the quantum state as a probability distribution of a quadrature observable depending on a range of lab parameters,
has been used to investigate quantum entanglement and failure of local realism, but has not to our knowledge led to many-mode
quantum simulations, presumably due to the lack of a positive propagator in nonlinear evolution. The complex P representation\cite{+P}
allows one to derive exact results for certain problems, but does not lead to stochastic equations, since the distribution
is neither real nor positive.

\section{Stochastic gauges}

The idea of stochastic gauges can be summarized for the generic case of any linear time-evolution problem which, like quantum
mechanics, can be expressed using a matrix product over a space of denumerable dimension.We first introduce a continuous
basis $\mathbf{\bm\Lambda}_{0}(\bm\alpha)$ in the underlying Hilbert space of operators, with unit trace: $\mathrm{Tr}[\mathbf{\bm\Lambda}_{0}(\bm\alpha)]=1$.
This must be an analytic function of the complex phase-space variables $\bm\alpha$, and have a mapping from the time-evolution
problem that generates only second-order phase-space derivatives. The basic mathematical steps are as follows:

\begin{enumerate}
\item \textbf{Wish to solve} $\,\partial\mathbf{\bm\rho/\partial}t=\mathbf{\mathbf{L}}\,\mathbf{\bm\rho}(t)\,\,\,$ where $\mathbf{\mathbf{L}}$
is a matrix and $\,\bm\rho$ is a vector of probabilities or amplitudes for the $d$ distinct occupation numbers $(N_{1},\dots,N_{d})$,
so its elements may be labeled $\rho_{N_{1},\dots,N_{d}}$. 
\item Introduce a $\,(d+1)$dimensional complex phase-space: $\,\underline{\bm\alpha}=(\,\Omega\,,\bm\alpha)$, with a renormalised
(gauge) basis of analytic vector functions $\mathbf{\bm\Lambda}(\underline{\bm\alpha})=\Omega\mathbf{\bm\Lambda}_{0}(\bm\alpha)$
\item Expand: $\,\mathbf{\bm\rho}=\int G(\underline{\bm\alpha})\mathbf{\bm\Lambda}(\underline{\bm\alpha})d^{2(d+1)}\underline{\bm\alpha}$
\item Equivalent time-evolution using second-order derivatives: \[
\,\frac{\partial}{\partial t}\mathbf{\bm\rho}(t)=\int G(\underline{\bm\alpha})\mathcal{L}_{A}\left[\mathbf{\bm\Lambda}(\underline{\bm\alpha})\right]d^{2(d+1)}\underline{\bm\alpha}\]

\item Add \emph{arbitrary} diffusion gauge vectors $\mathbf{f}(\underline{\bm\alpha})$ and drift gauges $\mathbf{g}(\underline{\bm\alpha})$
to give stability. 
\item \textbf{Equivalent stochastic equation:} $\,\partial\underline{\bm\alpha}\mathbf{/\partial}t=\underline{\mathbf{A}}+\underline{\underline{\mathbf{B}}}:\bm\zeta(t)$
\end{enumerate}

\subsection{Ladder Operators}

To explain the procedure in more detail, consider a generic equation, which is typically a type of master equation for a
quantum density matrix defined over a basis of number state occupation numbers: \begin{equation}
\,\frac{\partial}{\partial t}\mathbf{\bm\rho}(t)=\mathbf{\mathbf{L}}\,\mathbf{\bm\rho}(t)\,\,\,\end{equation}
 where $\,\mathbf{\bm\rho}$ has elements ${\rho_{N_{1},\dots,N_{d}}}$, with the labels $N_{j}$ corresponding to an occupation
of $N_{j}$ in mode $j$. If the density matrix has off-diagonal entries, these can be regarded as elements of an enlarged
vector, with $d=2M$ occupation numbers required for each entry in the case of $M$ modes. Note that this linear problem
is soluble in principle using diagonalization of $\mathbf{\mathbf{L}}$, but the (typically) large size of the matrix makes
this impractical.

Suppose $\mathbf{L}$ can be constructed from sums of products of matrix raising and lowering (`ladder') operators. These
either increase ($\,\mathbf{L}_{j}^{+}$) or decrease ($\,\mathbf{L}_{j}^{-}$) the number of particles and multiply the
probability by a function $\, f_{j}^{\pm}(N_{j})$. Thus, for master equations, one might have:\begin{equation}
\,\left[\mathbf{L}_{j}^{\pm}\,\mathbf{\bm\rho}\right]_{N_{1},\dots,N_{j},\dots,N_{d}}=f_{j}^{\pm}(N_{j})\rho_{N_{1},\dots,N_{j}\mp1,\dots,N_{d}}\end{equation}

This general structure occurs in many physical systems, including Pauli-type master equations for positive probabilities
found in many situations ranging from genetics to kinetic theory. In quantum problems, $\mathbf{\bm\rho}$ is a density matrix
in a number-state representation, while the ladder operators are bosonic creation and annihilation operators. The density
matrix can be written as an enlarged vector on a basis of number-state projectors with $d=2M$ for the case of bosonic many-body
theory on $M$ classical modes. In this case, $\mathbf{\bm\rho}(t)$ is not positive valued, but the time-evolution problem
is still linear.

There are also equations of identical structure found in 'imaginary time', which allow the calculation of canonical ensembles
in many-body theory. In these cases, the operator norm is not preserved, but the equations are still linear. To give a simple
example of the type of basis set that is of interest, a complete, analytic coherent state basis on the phase-space $\,\underline{\bm\alpha}=(\,\Omega\,,\bm\alpha)=(\,\Omega\,,\alpha,\beta)$
for the case of a single harmonic oscillator is given by:\begin{equation}
\mathbf{\bm\Lambda}(\underline{\bm\alpha})=\Omega|\alpha\rangle\langle\beta^{\ast}|e^{-\alpha\beta}\,\,,\label{cohstate}\end{equation}
 where $|\alpha\rangle$ is a bosonic coherent state.

\subsection{Identities}

Identities can now be constructed that depend on the nature of the continuous basis set $\bm\Lambda_{0}(\bm\alpha)$. We
only require that this basis set is an analytic function of the continuous variables $\bm\alpha$. While it is common to
use either Glauber coherent state projectors or Poisson distributions for this purpose, this is certainly not essential.
More general basis sets like $SU(N)$ coherent state projectors or general Gaussian bases are very likely to give even better
results, as they often more closely approximate the physical quantum states of interest.

Clearly, both raising and lowering identities are usually needed - in the following list, we indicate how the identities
map matrices onto differential operators in the phase space. This is just a generalised version of the well-known equivalence
between Heisenberg's matrix mechanics and Schroedinger's wave equation:\begin{eqnarray}
\mathbf{L}_{j}^{-}\,\mathbf{\bm\Lambda}(\underline{\bm\alpha}) & = & \mathcal{L}_{j}^{-}(\bm\partial,\bm\alpha)\left[\bm\Lambda(\underline{\bm\alpha})\right]\nonumber \\
\mathbf{L}_{j}^{+}\,\bm\Lambda(\underline{\bm\alpha}) & = & \mathcal{L}_{j}^{+}(\bm\partial,\bm\alpha)\left[\bm\Lambda(\underline{\bm\alpha})\right]\nonumber \\
\bm\Lambda(\underline{\bm\alpha}) & = & \Omega\partial_{\Omega}\left[\bm\Lambda(\underline{\bm\alpha})\right]\label{Ident}\end{eqnarray}

Provided $\,\bm\Lambda_{0}(\bm\alpha)$ is analytic in $\,\bm\alpha$, one can use $\,\underline{\bm\partial}=(\partial_{\Omega},\bm\partial)$
to symbolize either $\,\left[\partial_{j}^{x}\equiv\partial/\partial x_{j}\right]$ or $\,-i\left[\partial_{j}^{y}\equiv\partial/\partial y_{j}\right]$,
where $\,\alpha_{j}=x_{j}+iy_{j}$ for $j=0,1,\dots,d$, and $x_{j}$ as well as $y_{j}$ are real. These identities will
be used later to specify which form of the derivative will be used to obtain a positive-definite diffusion term.

\subsection{Diffusion gauge}

Using the identities to eliminate ladder operators, we obtain an evolution equation in integro-differential form: \begin{equation}
\,\frac{\partial}{\partial t}\mathbf{\bm\rho}(t)=\int G(\underline{\bm\alpha})\mathcal{L}_{A}\left[\mathbf{\bm\Lambda}(\bm\alpha)\right]d^{2(d+1)}\underline{\bm\alpha}\,\,\,.\end{equation}

Here the differential operator acts on the basis set, and must be of no more than second order:\begin{equation}
\,\mathcal{L}_{A}=U+A_{j}'\partial_{j}+\frac{1}{2}D_{ij}\partial_{i}\partial_{j}\,\,\,,\end{equation}
 where the implicit summation is over $i,j=1,\dots,d$ . At this stage, we notice that if we integrate by parts, we would
obtain a possible solution to the time-evolution provided that boundary terms vanish, and that:\begin{equation}
\,\frac{\partial}{\partial t}G(\underline{\bm\alpha},t)=\mathcal{L}_{N}\left[G(\underline{\bm\alpha})\right]\,\,\,.\end{equation}
 Here the normally ordered differential operator $\mathcal{L}_{N}$ is defined as:\begin{equation}
\,\mathcal{L}_{N}=U-\partial_{j}A_{j}'+\frac{1}{2}\partial_{i}\partial_{j}D_{ij}\,\,\,.\end{equation}

This type of generalised Fokker-Planck equation is known to be equivalent to a curved-space path integral\cite{Path-int}
on the complex phase-space. This means that we have indeed reduced the dimensionality of the problem, in the sense that the
phase-space dimensionality is much smaller than the dimensionality of the original vector. However, path-integrals are not
always convenient for numerical calculations, and we wish to transform the equations further into a stochastic form.

As the first step toward a stochastic reformulation, define a $\, d\times d'$ complex matrix square root $\,\mathbf{B}$
called the noise matrix, where: \begin{equation}
\,\mathbf{D}=\mathbf{BB}^{T}\end{equation}

Since this is non-unique, one can introduce diffusion gauges\cite{Plimak,GaugeP} from a set of matrix transformations $\mathbf{U}[\mathbf{f}(\underline{\bm\alpha})]$
with $\mathbf{UU}^{T}=\mathbf{I}$. Using these, it is clear that given an initial square root $\,\mathbf{B}'$, it is always
possible to construct another square root $\,\mathbf{B}$, corresponding to an alternative `diffusion gauge', with:\begin{equation}
\,\mathbf{B}=\mathbf{B}'\mathbf{U}[\mathbf{f}]\,\,\,.\end{equation}
 This is also an equally valid matrix square root, but it will in general have different stochastic properties.

The matrices $\mathbf{U}[\mathbf{f}]$ are just the usual set of complex orthogonal matrices where $\mathbf{U}[\mathbf{f}]\mathbf{U}^{T}[\mathbf{f}]=\mathbf{1}$.
These can either be fixed or variable functions of the phase-space coordinates, provided they satisfy growth restrictions.
If we assume that they are $d'\times d'$ square matrices, then they are generated by the antisymmetric $d'\times d'$ matrices,
that is, there are $d'\times(d'-1)/2$ diffusion gauges. Generally, $d'\leq d$, since the diffusion matrix can have zero
eigenvalues, although one can define larger noise matrices.

\subsection{Drift gauge }

A drift gauge Fokker-Planck equation is obtained as follows. Introduce $\: d'$ arbitrary complex functions $\,\mathbf{g}=(\, g_{j}(\underline{\bm\alpha},t)\,)$,
to give a new differential operator: \begin{equation}
\,\mathcal{L}_{G}=\mathcal{L}_{A}+\left[U+\frac{1}{2}\mathbf{g}\cdot\mathbf{g}\,\Omega\,\partial_{\Omega}+g_{k}B_{jk}\partial_{j}\right]\left[\Omega\partial_{\Omega}-1\right]\,\,\,.\end{equation}

Here, $\mathcal{L}_{G}$ \textbf{is equivalent} to $\,\mathcal{L}_{A}$, from the last identity in Eq (\ref{Ident}). Summing
indices over $i,j=0,\dots,d$ (where $i,j=0$ label the variable $\Omega$) , this can be rewritten in the form: \begin{equation}
\,\mathcal{L}_{G}=\left[A_{j}\partial_{j}+\frac{1}{2}D_{ij}\partial_{i}\partial_{j}\right]\,\,\,.\end{equation}
 This removes the non-stochastic term $\, U$, and - with the correct choice of gauge - stabilizes the drift equations.

\section{Stochastic gauge equations}

Since the initial drift vector was $\mathbf{A}'$, the \emph{total} complex drift vector, including gauge corrections is
$\,\underline{A}=(U\Omega,\mathbf{A})$, where:\begin{equation}
\,\mathbf{A}=\mathbf{A}'-\mathbf{B}\mathbf{g}\,\,.\end{equation}
 The total complex diffusion matrix $\,\underline{\underline{D}}$ is a $\,(d+1)\times(d+1)$ matrix, with a new $\,(d+1)\times d'$
square root $\,\underline{\underline{B}}$:\begin{equation}
\,\underline{\underline{D}}=\left[\begin{array}{cc}
\Omega^{2}\mathbf{g}\cdot\mathbf{g}, & \Omega\mathbf{g}^{T}\mathbf{B}^{T}\\
\mathbf{Bg}\Omega, & \mathbf{BB}^{T}\end{array}\right]=\left[\begin{array}{c}
\Omega\mathbf{g}^{T}\\
\mathbf{B}\end{array}\right]\left[\Omega\mathbf{g},\mathbf{B}^{T}\right]=\underline{\underline{B}}\,\underline{\underline{B}}^{T}\,\,\,.\end{equation}

\subsection{Dimension-doubling}

We now introduce the technique used to produce positive-definite diffusion, which depends on the analyticity of the basis
and the associated differential identities. This technique is identical to that used for the positive-P representation\cite{+P},
but is now extended to include the gauge amplitude variable as well. Define a $\,2(d+1)$ dimensional real phase space $\:(x_{0},..x_{d},y_{0}..\,,y_{d})$,
with derivatives $\,{\partial_{\mu}}$, where $\mu$ labels all the $2(d+1)$ real variables $x_{\mu}$.

Let: $\,\underline{\underline{\mathbf{B}}}=\underline{\underline{\mathbf{B}}}^{x}+i\underline{\underline{\mathbf{B}}}^{y}$
and $\,\underline{\mathbf{A}}=\underline{\mathbf{A}}^{x}+i\underline{\mathbf{A}}^{y}$ with all the ${}^{x}$ and ${}^{y}$
forms real. Choose the alternative analytic forms of the differential operator so that:

\begin{equation}
\,\underline{A}_{j}\partial_{j}\rightarrow\underline{A}_{j}^{x}\partial_{j}^{x}+\underline{A}_{j}^{y}\partial_{j}^{y}\,,\end{equation}

\begin{equation}
\,\underline{\underline{D}}_{ij}\partial_{i}\partial_{j}\rightarrow\underline{\underline{B}}_{ik}^{x}\underline{\underline{B}}_{jk}^{x}\partial_{i}^{x}\partial_{j}^{x}+\underline{\underline{B}}_{ik}^{y}\underline{\underline{B}}_{jk}^{x}\partial_{i}^{y}\partial_{j}^{x}+(x\leftrightarrow y)\,.\end{equation}
 With this identification of real derivatives, the original gauge differential operator is written:

\begin{equation}
\,\mathcal{L}_{G}=\left[\mathcal{A}_{\mu}\partial_{\mu}+\frac{1}{2}\mathcal{D}_{\mu\nu}\partial_{\mu}\partial_{\nu}\right]\,,\end{equation}

Next, on partial integration of the integral equation of motion, at least one valid solution for $G$ must satisfy: \begin{equation}
\frac{\partial}{\partial t}G=\left[-\partial_{\mu}\mathcal{A}_{\mu}+\frac{1}{2}\partial_{\mu}\partial_{\nu}\mathcal{D}_{\mu\nu}\right]G\end{equation}

By construction, the real diffusion matrix is a square of form:\begin{eqnarray}
\,\underline{\underline{\mathcal{D}}} & = & \left[\begin{array}{c}
\underline{\underline{\mathbf{B}}}^{x}\\
\underline{\underline{\mathbf{B}}}^{y}\end{array}\right]\left[(\underline{\underline{\mathbf{B}}}^{x})^{T},(\underline{\underline{\mathbf{B}}}^{y})^{T}\right]\\
 & = & \underline{\underline{\mathcal{B}}}\,\underline{\underline{\mathcal{B}}}^{T}\end{eqnarray}
 Clearly, $\underline{\underline{\mathcal{D}}}$ is positive definite. Hence, from the theory of stochastic equations\cite{Arnold},
provided some restrictions on growth are satisfied, one obtains the Ito stochastic differential equations:\begin{equation}
\,\frac{d}{dt}x_{\mu}=\mathcal{A}_{\mu}+\mathcal{B}_{\mu j}\zeta_{j}(t)\,\,,\end{equation}

where the real, Gaussian noise terms $\zeta_{j}(t)$ (for $j=1,\dots,d'$) are delta-correlated:

\begin{equation}
\langle\zeta_{i}(t)\zeta_{j}(t')\rangle=\delta(t-t')\delta_{ij}\,\,.\end{equation}

\subsection{Central result of stochastic gauge theory}

Another, clearer way to write this result is to return to a complex vector notation. Ito stochastic equations for the complex
trajectory and gauge amplitude $\Omega$ are therefore obtained as follows:\begin{eqnarray}
\frac{d\Omega}{dt} & = & \Omega\left[U+\,\mathbf{g}\cdot\bm\zeta(t)\right]\nonumber \\
\frac{d\bm\alpha}{dt} & = & \mathbf{A}'+\mathbf{B}:\left[\bm\zeta(t)-\,\mathbf{g}\right]\label{Central}\end{eqnarray}

Here, the arbitrary gauge terms $\,\mathbf{g}$ can be used to eliminate moving singularities that might be already present
with the analytic drift $\mathbf{A}'$; the actual simulated drift is $\mathbf{A}=\mathbf{A}'-\mathbf{B}:\,\mathbf{g}$.
Gauges can be chosen freely to optimize simulations, in either real or imaginary time.

It is essential to recognize that only the basis set, not the gauges, must be analytic functions on the phase space. Thus,
while the original drift is usually analytic, the gauge modified drift is best chosen not to be analytic. This is because
the analytic continuation of systems of (generically) non-integrable nonlinear equations typically has moving singularities,
which are trajectories that can deterministically reach infinity in a finite time. This is related to the Painleve conjecture
of mathematical physics\cite{Painleve}. To remove the singularities, with their resulting boundary terms, a non-analytic
gauge is therefore needed. Provided there are no boundary terms, all gauges are the same physically - but in practice, they
give rise to different sampling errors, and therefore must be optimized carefully.

Although removal of moving singularities appears necessary, this is not sufficient either to minimize sampling error or to
guarantee the absence of boundary terms. In general, gauges must still be checked on a case by case basis. We have found
that diffusion gauges that have no radial growth in the extended phase-space, and drift gauges where all phase-space trajectories
are directed toward the origin at large enough radius, appear to eliminate boundary terms in most physically sensible examples.
Clearly, a more rigorous investigation into these issues is still needed, since there may be anomalies even with these restrictions.

\subsection{Observables}

We typically wish to calculate physically observable quantities or moments of $\mathbf{\mathbf{\bm\rho}}$ in the form of
a trace, where for this purpose the quantum density operator should be regarded as a matrix:\begin{eqnarray*}
\langle\mathbf{O}\rangle & = & \frac{\mathrm{Tr}[\mathbf{O\mathbf{\bm\rho}}]}{\mathrm{Tr}[\mathbf{\bm\rho}]}\\
 & = & \frac{\int G(\underline{\bm\alpha})\mathrm{Tr}[\mathbf{O\mathbf{\bm\Lambda}}(\underline{\bm\alpha})]d^{2(d+1)}\underline{\bm\alpha}}{\int G(\underline{\bm\alpha})\Omega d^{2(d+1)}\underline{\bm\alpha}}\end{eqnarray*}

If the problem involves Bose operators, with a coherent state basis, then the Hermitian observables can be written in a normally
ordered form, $\widehat{O}=O_{N}(\mathbf{a},\mathbf{a}^{\dagger})$. The equivalent c-number expression for ensemble averaging
is:

\begin{equation}
O(\underline{\bm\alpha})=\Omega O_{N}(\bm{\alpha},\bm{\beta})\,\,,\label{Observable}\end{equation}
 so the quantum ensemble average $\left\langle \widehat{O}\right\rangle $has an immediate expression as: \begin{equation}
\left\langle \widehat{O}\right\rangle =\int\frac{G(\underline{\bm\alpha})}{\mathcal{N}}O(\underline{\bm\alpha})d^{4M+2}\underline{\bm\alpha}=\frac{\left\langle O(\underline{\bm\alpha})\right\rangle _{S}}{\mathcal{N}}\,.\label{expectation}\end{equation}

Here $\mathcal{N}=\left\langle \Omega\right\rangle _{S}$ , to preserve the trace of the normalized density matrix, and $\left\langle O\right\rangle _{S}$
represents a stochastic average on the phase-space of all trajectories $\underline{\bm\alpha}$ , including the weighting
factor $\Omega$ at each point in the trajectory.

\section{Examples}

The focus of this section is to give examples in simple cases that are exactly soluble, yet with statistics that are far
from Poissonian. The reason for this is to illustrate how the stochastic gauge technique is successful in treating cases
that would not be soluble using any previous simulation method, since the statistics are quite different to those of the
underlying basis set. As examples, we will start with a simple chemical reaction master equation in which there are no quantum
coherences, then move to a canonical ensemble example, and finally a quantum dynamics problem.

None of the examples presents any real difficulties, since they are exactly soluble. However, our purpose here is to show
that the stochastic gauge method gives correct results in cases where the solutions are already known. This, of course, is
an essential first step toward treating more complex systems where the results are not known a priori. It is also interesting
to see how these techniques have rather general applicability in physics and related scientific fields. This allows the possibility,
for example, of combining imaginary time propagation for the initialization of the quantum system in a thermal ensemble,
followed by real time propagation to simulate the response to a change in the Hamiltonian. Such experiments are common in
many-body physics.

\subsection{Master equations}

The first example will be a type of Pauli master equation, in which there are only diagonal terms in the density matrix -
so there are only simple probabilities in the original equations. These types of equation commonly occur in chemical\cite{Poisson}
and biological\cite{eco} applications. A typical example is the astrophysically important problem of the formation of molecular
hydrogen on interstellar grain surfaces\cite{Biham}. A simplified reaction model is then:\begin{eqnarray*}
H^{(IN)} & :\rightarrow_{r} & H\\
2H & \rightarrow_{k} & H_{2}\\
H & \rightarrow_{\gamma} & H^{*}\end{eqnarray*}

This describes adsorption of hydrogen atoms via a rate ($r$) from an input flux $H^{(IN)}$ , together with desorption at
a rate $\gamma$. In addition molecule formation occurs at a rate of $k$. The corresponding master equation can be transformed
to Fokker-Planck form using the Poisson representation, giving an analytic steady-state solution for the $m-$th moment of
$n$ (the number of $H$ atoms), in terms of Bessel functions:\begin{eqnarray}
\left\langle n^{m}\right\rangle  & =C & \int_{-\infty}^{(0+)}\mu^{(2-m-\gamma/k)}e^{r\mu/k+2/\mu}d\mu\nonumber \\
\nonumber \\ & = & \left(\frac{r}{2k}\right)^{m/2}I_{\gamma/k+m-1}(4\sqrt{r/2k})/I_{\gamma/k-1}(4\sqrt{r/2k})\,\,\,.\label{eq:moment}\end{eqnarray}
 This gives the $H_{2}$ production rate via $R_{H_{2}}=k\left\langle n^{2}\right\rangle $ .

\subsubsection{Stochastic equations}

Here we use an exact expansion of the distribution vector $\bm\rho$ using `prototype' solutions, namely the complex Poisson
distribution $\bm\Lambda_{0}(\bm\alpha)$:\begin{equation}
\left[\bm\Lambda_{0}(\bm\alpha)\right]_{N_{1},\dots,N_{d}}=\prod_{\mathbf{j}=1}^{d}e^{-\alpha_{j}}\left(\alpha_{j}\right)^{N_{j}}/N_{j}!\,\,\,\end{equation}
 By comparison, the original Poisson representation\cite{Poisson} of Gardiner expands the distribution vector with a positive
distribution of Poissonians, $f(\bm\alpha)$, defined over a \emph{complex} $d$-dimensional phase-space of variables $\bm\alpha$,
just as we do here - but without the extra weight-factor $\Omega$. Including the weight factor, we can take advantage of
the more general stochastic gauge procedure summarized in Eq (\ref{Central}). Together with the corresponding Poisson identities,
one finds the following Ito stochastic equations for molecule production including the gauge terms, where $\langle\zeta(\tau)\zeta(\tau')\rangle=\delta(\tau-\tau')$:

\begin{eqnarray}
\frac{d\Omega}{d\tau} & = & \Omega g\zeta\,\nonumber \\
\frac{dn}{d\tau} & = & \left[r-\gamma n-2kn^{2}\right]+i\sqrt{2k}n[\zeta-g]\,\,\,.\label{eq:Ito}\end{eqnarray}

If there is no gauge, the result of Fig (1) is obtained, corresponding to the original Poisson representation method.

\begin{figure}
\begin{center}\includegraphics[%
  width=8cm,
  keepaspectratio]{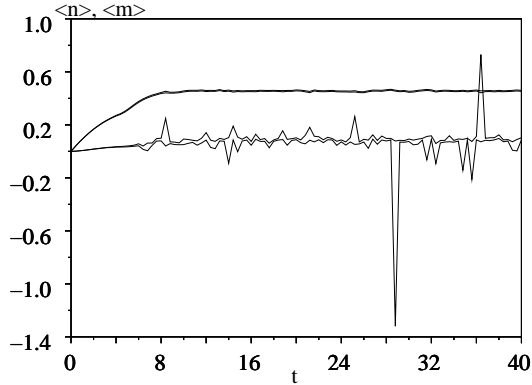}\end{center}

\caption{Sampled moments of $\langle n\rangle$ (upper curve) and $\langle n^{2}\rangle$ (lower curve) for astrophysical hydrogen
molecule production in the Poisson representation, parameters as in text. Adjacent lines give upper and lower error bars
caused by sampling error. The numerical values used here were $k=0.5$, $\gamma=0.1$, $r=0.1$ .}
\end{figure}

This result is clearly extremely inaccurate. It has a large sampling error in $\langle n^{2}\rangle$ and we will see that
it also has a systematic error in $\langle n\rangle$. The reason for this is that the original ungauged equations have an
instability as $n\rightarrow-\infty$, leading to a moving singularity. This causes power-law tails and systematic boundary
term errors in the resulting phase-space distribution, when there are small $\gamma/k$ ratios.

Fortunately, it is simple to stabilize these equations by adding non-analytic corrections to the drift. The simplest case
is the `circular' gauge, which replaces the analytic variable $n$ by its modulus $|n|$:\[
g_{c}=i\sqrt{2k}(n-|n|)\]

In this gauge, the Ito equations are:\begin{eqnarray}
\frac{d\Omega}{d\tau} & = & i\Omega(n-|n|)\,\sqrt{2k}\,\zeta\nonumber \\
\frac{dn}{d\tau} & = & r-n\left[\gamma+2k|n|\right]+in\sqrt{2k}\zeta\,\,\,.\label{eq:Strat}\end{eqnarray}
 The corresponding results are shown in Fig (2), indicating a dramatically improved sampling error, and no systematic errors.

\begin{figure}
\begin{center}\includegraphics[%
  width=8cm,
  keepaspectratio]{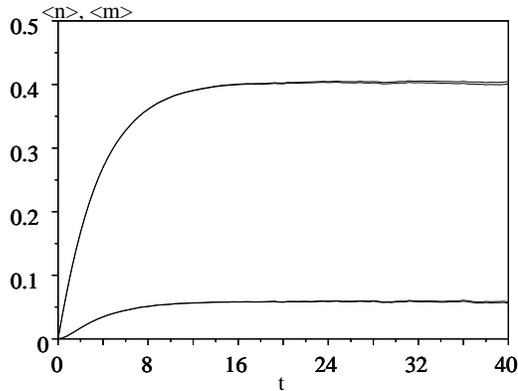}\end{center}

\caption{Sampled moments of $\langle n\rangle$ (upper curve) and $\langle n^{2}\rangle$ (lower curve) for astrophysical hydrogen
molecule production in the `circular' gauge, parameters as in text. Adjacent lines give upper and lower error bars caused
by sampling error, which is invisible on this scale.}
\end{figure}

For the circular gauge and for the Poisson expansion, the observed moment and its sampling error is given in Table 2, which
tabulates the final near-equilibrium simulation results at $t=40$, and compares them to the equilibrium analytic result
for $t=\infty$. For the stable circular gauge, the results are within one standard deviation of the analytic calculation
in all cases. Other gauges are also possible - in fact, almost any gauge which suppresses the moving singularities will give
acceptable results, as long as no new boundary terms are introduced by the gauge itself.

\begin{table}
\begin{center}\begin{tabular}{|c|c|c|c|}
\hline 
Moment&
Analytic&
Poisson&
Circular gauge\tabularnewline
\hline
\hline 
$\langle\Omega\rangle$&
$1.0$&
$1.0$&
$.994(10)$\tabularnewline
\hline 
$\langle n\rangle$&
$0.407..$&
$0.457(4)$&
$0.402(5)$\tabularnewline
\hline 
$\langle n^{2}\rangle$&
$0.059..$&
$0.094(8)$&
$0.061(2)$\tabularnewline
\hline
\end{tabular}\end{center}

\caption{Table of observed moments, comparing analytic results with those for the Poisson representation and for the `circular' stochastic
gauge. The moment $\langle n^{2}\rangle$ is critical for molecule production. Sampling error in brackets.}
\end{table}

By comparison, the unstable ungauged Poisson method clearly gives enormous sampling errors, with incorrect averages in $\langle n\rangle,$
due to systematic boundary term errors. While this problem is relatively simple (for purposes of illustration), the stochastic
techniques given here are easily extended to more complex problems where the original master equations cannot be solved directly.
More details of this will be given elsewhere\cite{Gaugepoisson}.

\subsection{Canonical ensemble}

For computational purposes, we can reduce the Bose gas Hamiltonian to a lattice Hamiltonian which contains all the essential
features. This includes nonlinear interactions at each of $M$ sites or modes, together with linear interactions that couple
different sites together. Such problems are important for quantum gases trapped in optical lattices, or in low-dimensional
environments, where evidence for quantum coherence and particle antibunching has been inferred in recent experiments. 

The simplest case that can represent a Bose-Einstein condensate (BEC) in a one-mode trap has $M=1$, so:

\begin{equation}
\widehat{H}=\hbar\omega:\widehat{n}:+\hbar\chi:\widehat{n}^{2}:\,.\label{Hamiltonian}\end{equation}

In this normally ordered Hamiltonian, the operator $\widehat{n}=\widehat{a}{}^{\dagger}\widehat{a}$ is the boson number
operator. The above Hamiltonian can be easily generalized to many important interacting Bose gas models. The canonical ensemble
in thermal equilibrium for the one-mode case is an exactly soluble problem, which can be used to illustrate the gauge method.
Applications in less trivial cases will appear elsewhere. It is an interesting historical note that the quantum correction
to a classical canonical ensemble calculation was the first application\cite{Wig-Wigner} of the Wigner distribution, and
Wigner regarded the effect of Bose or Fermi statistics to be a serious issue to be addressed in future.

The un-normalized grand canonical quantum density matrix is defined for our purposes as a slight modification to the usual
form in statistical mechanics. We let: \begin{equation}
\widehat{\rho}=\exp\left[-\tau\widehat{K}-\varepsilon\widehat{N}\right]\,\,,\label{rho-defn}\end{equation}
 where $\tau=\hbar/k_{B}T$, and $\widehat{K}(\mu,\widehat{a},\widehat{a}^{\dagger})=\widehat{H}/\hbar-\mu\widehat{N}$.
We choose $\varepsilon\ll1$ to give a high-temperature initial state at $\tau=0,$ with an initial occupation of $n_{0}=1/\left[\exp(\varepsilon)-1\right]\approx1/\varepsilon$
at each site. Thus, the effective chemical potential is actually $\mu_{\mathrm{eff}}=\mu-\varepsilon/\tau$. Since $\left[\widehat{K},\widehat{N}\right]=0$,
direct differentiation of Eq (\ref{rho-defn}) shows that the density matrix satisfies the equation:\begin{equation}
\frac{\partial}{\partial\tau}\widehat{\rho}(\tau)=-\frac{1}{2}\left[\widehat{K},\widehat{\rho}\right]_{+\,\,.}\label{rho-evolve}\end{equation}
 Solving this equation with the initial conditions at $\tau=0$ gives the solution for $\widehat{\rho}(\tau)$ at lower temperatures,
where quantum effects like Bose condensation will occur.

Let us expand the density matrix $\widehat{\rho}$ on an off-diagonal coherent state basis set in the manner of the positive
P distribution. This is given in (\ref{cohstate}). The initial G-distribution is Gaussian: \begin{equation}
G_{0}(\underline{\bm\alpha})\propto\exp\left[-\left|\alpha\right|^{2}/n_{0}\right]\delta^{2}(\alpha-\beta^{*})\delta^{2}(\Omega-1)\,\,.\label{P0}\end{equation}

To determine the effects of the `Kamiltonian' $\widehat{K}$ on $G(\underline{\bm\alpha})$, it is first necessary to calculate
the effect of the annihilation and creation operators on the projectors $\bm\Lambda(\underline{\bm\alpha})$. This is obtained
as follows: \begin{eqnarray}
\widehat{a}\,\bm\Lambda(\underline{\bm\alpha}) & = & \alpha\,\bm\Lambda(\underline{\bm\alpha})\nonumber \\
\widehat{a}^{\dagger}\bm\Lambda(\underline{\bm\alpha}) & = & \left[\partial_{\alpha}+\beta\right]\bm\Lambda(\underline{\bm\alpha})\nonumber \\
\bm\Lambda(\underline{\bm\alpha}) & = & \Omega\partial_{\Omega}\bm\Lambda(\underline{\bm\alpha})\end{eqnarray}
 together with the corresponding identities for the reversed orderings. Using these operator identities, the operator equation
(\ref{rho-evolve}) can be transformed to a differential equation. The (ungauged) differential operator acting on the basis
$\bm\Lambda(\underline{\bm\alpha})$ is \begin{eqnarray}
\mathcal{L}_{A} & = & -\frac{1}{2}\left[K(\mu,\alpha,\partial_{\alpha}+\beta)+K(\mu,\partial_{\beta}+\alpha,\beta)\right]\\
 & = & -K(\mu,\alpha,\beta)+\sum_{j=1}^{2}\left[A'_{j}\partial_{j}+\frac{1}{2}D_{j}\partial_{j}^{2}\right]\,.\end{eqnarray}
 To simplify notation, define $n=\alpha\beta$, $\alpha_{1}=\alpha$, and $\alpha_{2}=\beta$. Then $A'_{j}=(\mu-2\chi n-\omega)\alpha_{j}/2$
, and $D_{j}=-\chi\alpha_{j}^{2}$ . 

Following the main procedure summarized in Eq (\ref{Central}), a gauge correction is utilized to stabilize coherent state
paths in highly non-classical regions in phase-space. This allows one to benefit greatly from the over-completeness of coherent
states, in reducing the sampling error and eliminating boundary terms. To stabilize large modulus trajectories which otherwise
can lead to boundary term errors and large sampling uncertainties, we choose drift gauges $g_{j}=i\sqrt{\chi}(n-|n|)$, giving:\begin{eqnarray}
\frac{d\alpha_{j}}{d\tau} & = & \left[\left(\mu-\omega\right)/2-\chi\left|n\right|\right]\alpha_{j}+i\alpha_{j}\sqrt{\chi}\zeta_{j}\nonumber \\
\frac{d\Omega}{d\tau} & = & \left[-K(\mu,{\bm\alpha})+\sum_{j=1}^{2}g_{j}\zeta_{j}\right]\Omega\end{eqnarray}

There is an intuitive physical interpretation. Since $\beta=\alpha^{*}$ in the initial thermal state, each amplitude initially
obeys a Gross-Pitaevskii equation in imaginary time, with quantum phase-noise due to the interactions. This causes non-classical
statistics with ${\alpha}\neq{\beta}^{*}$ to emerge at low temperatures. Along each path an additional ensemble weight $\Omega$
is accumulated, which is logarithmically proportional to the Kamiltonian $K({\bm\alpha})$. The zero-temperature steady-state
is the usual Gross-Pitaevskii approximation, together with quantum corrections.

To illustrate the method, first consider the non-interacting case with $\chi=0$, where we can set the gauge to zero, and
define $\alpha=\beta^{\ast}$ , giving a diagonal Glauber P-distribution. Then:\begin{eqnarray}
\frac{dn}{d\tau} & = & (\mu-\omega)n\nonumber \\
\frac{d\Omega}{d\tau} & = & \Omega(\mu-\omega)n\label{free}\end{eqnarray}

The power of the normal-ordered coherent state expansion is shown by the fact that these equations are deterministic, \emph{even
though they include all quantum fluctuations}. By contrast, the corresponding path-integral equations have large vacuum noise
terms. These equations can be integrated immediately to give a Bose-Einstein thermal ensemble with Gaussian fluctuations
: \begin{equation}
\left\langle \widehat{n}\right\rangle _{\tau}=\frac{\left\langle n(\tau)\Omega(\tau)]\right\rangle _{S}}{\left\langle \Omega(\tau)\right\rangle _{S}}=\frac{1}{\exp([\omega-\mu_{\mathrm{eff}}]\tau)-1}\,\,.\label{thermal}\end{equation}

Next, consider the exactly soluble interacting case, involving a single Bose mode with:\begin{equation}
\widehat{H}(a,a^{\dagger})=\hbar\chi:\widehat{n}^{2}:\,.\label{anharmonic}\end{equation}

\begin{center}%
\begin{figure}
\begin{center}\includegraphics[%
  width=8cm,
  height=8cm]{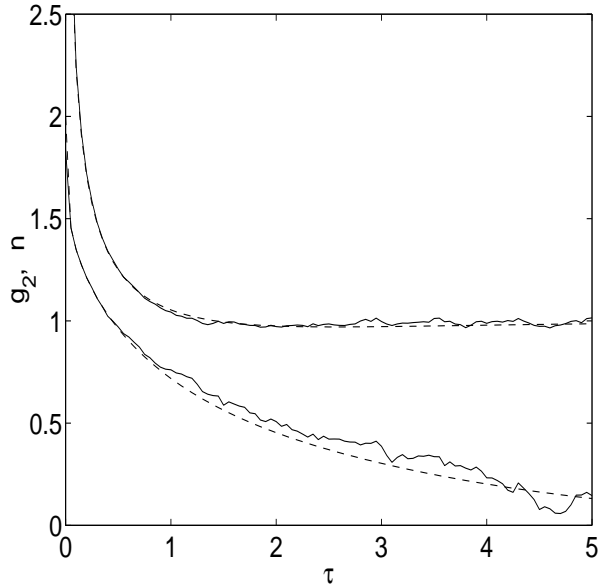}\label{figg2nbar}\end{center}

\caption{Simulation (solid line) versus exact results (dotted line) for the boson number $\mathrm{n}=\langle\widehat{n}\rangle$
and correlation function $\mathrm{g}_{2}=\langle:\!\widehat{n}^{2}\!:\rangle/\mathrm{n}^{2}$ of the exactly soluble anharmonic
oscillator case with parameters $\chi=\mu=0.5$, $\varepsilon=0.1$\label{g2nbar}.}
\end{figure}
\end{center}

A numerical simulation is of most interest here, as it can be generalized to other Bose gas systems of greater complexity.
It is straightforward to obtain agreement with the exact solution for large boson number, as quantum-noise corrections are
small in this limit. Instead, we focus on the case furthest from coherent statistics with $\mu=\chi=0.5$, giving just one
boson in the zero-temperature limit, and choose $\varepsilon=0.1$. This case is shown in Fig (\ref{figg2nbar}), as well
as a comparison with the exact results.

\vspace{0.3cm}

\vspace{0.3cm}

The results can be seen to agree well with the exact analytic ones, and the agreement is easily improved by increasing the
number of stochastic trajectories. This excellent agreement also occurs with much larger numbers of bosons. The physical
behaviour of strong particle antibunching is in agreement with evidence deduced from recent experiments in optical lattices.

\subsection{Anharmonic oscillator}

\label{anharm}

The quantum dynamics of the anharmonic oscillator is the subject of much current interest. One can combine the previous canonical
technique with a real-time evolution, in order to model a transient experiment in which a BEC is first cooled, then allowed
to evolve after a change in the Hamiltonian. We consider the Hamiltonian to be Eq (\ref{Hamiltonian}), the same as previously.

This type of problem has been studied previously as part of more extended multi-mode studies on quantum solitons, leading
to the prediction and observation of quantum squeezing in solitons\cite{carterdrummond:87,qsol2}. Other applications include
first-principles studies of evaporative cooling\cite{evapcool}, and a treatment of phase-diffusion\cite{becpf} using an
approximate Wigner technique, starting from a coherent state. The present technique is exact rather than approximate, though
it predicts, as expected, the same behaviour of phase-diffusion and amplitude decay from an initial coherent state. However,
we shall just consider the one-mode case here.

We find the following salient points:

\begin{itemize}
\item Diffusion gauges work better than drift gauges at controlling sampling error 
\item Sampling error grows with time 
\item Diffusion gauges work by trading off (increasing) phase noise against intensity noise 
\item Even better results occur if drift and diffusion gauges are combined 
\item It is possible to simulate past the time of amplitude decay 
\end{itemize}
As previously, we use the stochastic gauge procedure of Eq (\ref{Central}). The resulting Ito equations in an arbitrary
drift gauge are as follows, where $\alpha,$ $\beta$ are the two variables that correspond to $\hat{a},\hat{a}^{\dagger}$
in a coherent state (positive-P) expansion. We define $\tau=\chi t$, $\omega=0$, and $n=\alpha\beta$, corresponding to
$\widehat{n}=\hat{a}^{\dagger}\hat{a}$:

\begin{eqnarray}
\frac{d\alpha}{d\tau} & = & -2in\alpha+(1-i)\alpha(\zeta_{1}-g_{1})\nonumber \\
\frac{d\beta}{d\tau} & = & 2in\beta+(1+i)\beta(\zeta_{2}-g_{2})\label{Alphaequn}\\
\frac{d\Omega}{d\tau} & = & \sum_{j=1}^{2}g_{j}\zeta_{j}\Omega\,\,.\nonumber \end{eqnarray}
 Here $\langle\zeta_{i}(\tau)\zeta_{j}(\tau')\rangle=\delta_{ij}\delta(\tau-\tau')$.

These equations correspond to a diagonal noise matrix of form:

\begin{equation}
\mathbf{B}'=\left[\begin{array}{cc}
(1-i)\alpha, & 0\\
0, & (1+i)\beta\end{array}\right].\label{Bmat}\end{equation}
 It is convenient to use an equivalent diffusion gauge with a noise matrix $\mathbf{B}=\mathbf{B}'\mathbf{U}$ defined in
terms of a parameter $A$, using an orthogonal transform $\mathbf{U}$ so that:

\begin{equation}
\mathbf{U}=\left[\begin{array}{cc}
\cosh A, & -i\sinh A\\
i\sinh A, & \cosh A\end{array}\right]\,\,.\label{Umat}\end{equation}
 We also introduce new variables $\theta,\phi$ where:\begin{eqnarray}
n & = & e^{\theta}\nonumber \\
\alpha/\beta & = & e^{i\phi}\,\,.\label{ntheta}\end{eqnarray}

These variables are interpreted as the logarithmic amplitude and phase respectively. Their equations including gauge terms
are:

\begin{eqnarray}
\frac{d\phi}{d\tau} & = & 2-4n-e^{A}(1+i)\left[(\zeta_{1}-g_{1})-i(\zeta_{2}-g_{2})\right]\nonumber \\
\frac{d\theta}{d\tau} & = & e^{-A}(1-i)\left[(\zeta_{1}-g_{1})+i(\zeta_{2}-g_{2})\right]\label{phiequn}\\
\frac{d\Omega}{d\tau} & = & \sum_{j=1}^{2}g_{j}\zeta_{j}\Omega\nonumber \end{eqnarray}

Without any drift gauge, these equations have the problem that the logarithmic ratio of amplitudes (imaginary phase) $|\Im[\phi]|$
can grow rapidly whenever the quantum noise causes $n$ to have an imaginary part, resulting in a large sampling error. This
can be controlled to some extent with the diffusion gauge technique\cite{Plimak}, which allows us to make $A$ large and
positive, thereby reducing the quantum noise in $n$. This, however, is at the expense of an increase in quantum noise in
the phase, and can only delay the onset of the rapid growth in $\Im[\phi]$.

\begin{figure}
\begin{center}\includegraphics[%
  width=8cm,
  keepaspectratio]{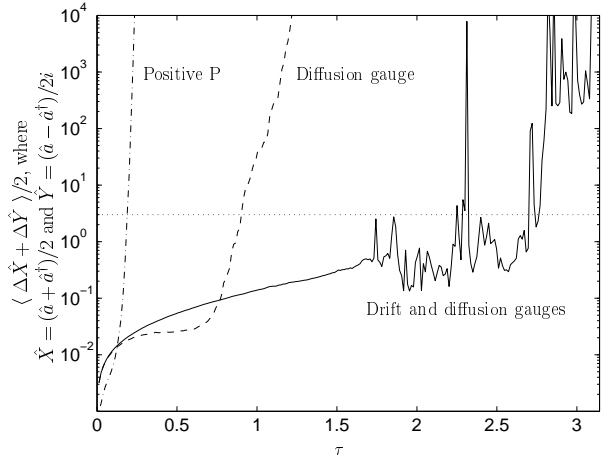}\end{center}

\caption{Graph of mean sampling error in expectation values of quadrature moments versus time, using several different types of gauge.
The combined diffusion and drift gauges give the best result, with an order of magnitude longer time duration before the
sampling error becomes substantially large. Coherent state initial conditions with amplitude $\alpha(0)=3$. Constant diffusion
gauge $A=1.4$, $10^{5}$ trajectories.\label{Graph-of-sampling-error}}
\end{figure}

Even better results are obtained by combining diffusion and drift gauges. We choose the gauge $g_{1}=(1+i)\Im[n]e^{-A}=-ig_{2}$.
This has the property that only noise drives the ratio of amplitudes\begin{equation}
\frac{d\Im[\phi]}{d\tau}=e^{A}(\zeta_{2}-\zeta_{1})\label{gaugeequn}\end{equation}
 as opposed to the un-gauged equation of:\begin{equation}
\frac{d\Im[\phi]}{d\tau}=-4\Im[n]+e^{A}(\zeta_{2}-\zeta_{1})\label{nogauge}\end{equation}

Typical results are shown in Figure (\ref{Graph-of-sampling-error}), which gives the time-dependence of the sampling error
with different gauges. This indicates substantial increases in useful simulation times compared to any previous phase-space
technique\cite{ccp2k,Plimak}. We note that the sampling error still increases with time. This appears to be inevitable with
current stochastic gauge methods, which stabilize trajectories but introduce an increasing uncertainty in the relevant quantum
amplitude.

\section{Summary}

Complexity is a major, even central problem in modern theoretical physics. Our fundamental description of nature, quantum
field theory, is incredibly complex in terms of the Hilbert space dimension - far more so than any classical description.
Mappings to phase-spaces of reduced dimensionality therefore provide an attractive route to overcoming this complexity problem.
For best results, however, one wishes to have a stochastic mapping, where the dynamics in phase-space obey a local stochastic
description. In these cases the mapping permits a way to sample the complex dynamics over a finite set of samples, thus providing
a controllable approximation to the exact dynamics.

The Wigner representation is the pioneering method in phase-space that maps quantum mechanics into a classical phase-space.
However, it has some drawbacks. It gives large vacuum fluctuations in quantum field simulations, and is essentially non-stochastic,
as it is not a positive-definite representation. All the other techniques that have been introduced for classical phase-spaces,
like the Glauber P-representation, have similar problems. This includes even the Q-representation - which is always positive,
but has no local positive propagator when the Hamiltonian is nonlinear. More modern techniques like the positive-P representation,
solve most of the technical problems due to lack of positivity, but can give boundary term errors due to moving singularities
in the drift equations.

Gauge techniques solve these known mathematical problems by stabilizing the drift equations. However, in specific cases,
one still needs to find the optimum method or gauge field. Several examples of workable gauges in different cases and bases
have been given here, and their utility demonstrated by comparison with exact results. In the long run, such techniques can
allow one to make progress toward calculations that involve interacting many-body systems. We note that the gauge approach
is rather general, since in any particular case one can optimize the basis set, the gauge, and even the algorithm for sampling
the weighted trajectories.

From the larger theoretical perspective these are important issues, as we treat more challenging complex systems, including
possible tests of quantum mechanics in new regions of macroscopic and entangled quantum systems.


\begin{thebibliography}{10}
\bibitem{feynman:82}R.~P.~Feynman, Int.~J.~Theor.~Phys. \textbf{21}, 467 (1982). 
\bibitem{GaugeP}P. Deuar and P. D. Drummond, Phys. Rev. A \textbf{66}, 033812 (2002). 
\bibitem{Wig-Wigner}E.~P.~Wigner, Phys.~Rev. \textbf{40}, 749 (1932). 
\bibitem{wilson:74}K.~G.~Wilson, Phys.~Rev.~D \textbf{10}, 2445 (1974); M. Creutz, Phys. Rev. D \textbf{21}, 2308 (1980). 
\bibitem{Ceperley}E. L. Pollock and D. M. Ceperley,~Phys. Rev. B \textbf{30}, 2555 (1984); D.~M.~Ceperley, Rev.~Mod.~Phys. \textbf{67},
279 (1995); D.~M.~Ceperley, Rev.~Mod.~Phys. \textbf{71} S438 (1999). 
\bibitem{Hus-Q}K.~Husimi, Proc.~Phys.~Math.~Soc.~Jpn. \textbf{22}, 264 (1940). 
\bibitem{Gla-P}R.~J.~Glauber, Phys.~Rev. \textbf{131}, 2766 (1963); K.~E.~Cahill and R.~J.~Glauber, Phys.~Rev. \textbf{177}, 1882
(1969). 
\bibitem{Sud-P}E.~C.~G.~Sudarshan, Phys.~Rev.~Lett. \textbf{10}, 277 (1963). 
\bibitem{+P}S. Chaturvedi, P. D. Drummond and D. F. Walls, J. Phys. A \textbf{10}, L187-192 (1977); P.~D.~Drummond and C.~W.~Gardiner,
J.~Phys.~A \textbf{13}, 2353 (1980). 
\bibitem{SG-abs}A.~M.~Smith and C.~W.~Gardiner, Phys.~Rev.~A \textbf{39}, 3511 (1989). 
\bibitem{SS-fail}R.~Schack and A.~Schenzle, Phys.~Rev.~A \textbf{44}, 682 (1991). 
\bibitem{GGD-Validity}A.~Gilchrist, C.~W.~Gardiner, and P.~D.~Drummond, Phys.~Rev.~A \textbf{55}, 3014 (1997). 
\bibitem{Paris1}I. Carusotto, Y. Castin and J. Dalibard,~Phys. Rev. A \textbf{63}, 023606 (2001). 
\bibitem{ccp2k}P.~Deuar and P.~D.~Drummond, Comp.~Phys.~Comm. \textbf{142}, 442 (2001). 
\bibitem{Paris2}I. Carusotto, Y. Castin, J. Phys. B \textbf{34}, 4589 (2001). 
\bibitem{Plimak}L. I. Plimak, M. K. Olsen, M. J. Collett, Phys. Rev. A \textbf{64}, 025801 (2001). 
\bibitem{Mancini}S.~Mancini, V.~I.~Man'ko, and P.~Tombesi, Phys.~Lett.~A \textbf{213}, 1 (1996); Found.~Phys. \textbf{27}, 801 (1997).
\bibitem{Painleve}T. Bountis, H. Segur and F. Vivaldi, Phys. Rev. A \textbf{25}, 1257 (1982). 
\bibitem{Path-int}R. Graham, Z. Physik B \textbf{26}, 281 (1977); H. Leschke and M. Schmutz, Z. Physik B \textbf{27}, 85 (1977). 
\bibitem{Arnold}L. Arnold, \emph{Stochastic Differential Equations: Theory and Applications} (Wiley, New York, 1974); C. W. Gardiner, \emph{Handbook
of Stochastic Methods} (Second Edition, Springer, Berlin, 1985). 
\bibitem{Poisson}I. Matheson, D. F. Walls, C. W. Gardiner, J. Stat. Phys \textbf{12}, 21 (1975); C.W. Gardiner, S. Chaturvedi, J. Stat. Phys.
\textbf{17}, 429 (1977); 18, 501 (1978); C. W. Gardiner, \emph{Handbook of Stochastic Methods}, (2nd Ed, Springer, Berlin,
1985). 
\bibitem{eco}M. Eigen, Naturwissenschaften \textbf{58}, 465 (1971); I. Hanski, Nature \textbf{396}, 41 (1998); D. Alves and J. F. Fontanari,
Phys. Rev E \textbf{57}, 7008 (1998); B. Drossel, Advances in Physics \textbf{50}, 209 (2001). 
\bibitem{Biham}S. B. Charnley, Astrophys J 509, \textbf{L121} (1998); Astrophys J 562, \textbf{L99} (2001);.O. Biham, I. Furman, V. Pirronello
and G. Vidali, Astrophys. J. \textbf{553}, 595~(2001). 
\bibitem{Gaugepoisson}P. D. Drummond, to be published. 
\bibitem{carterdrummond:87}S.~J.~Carter, P.~D.~Drummond, M.~D.~Reid, and R.~M.~Shelby, Phys.~Rev.~Lett. \textbf{58}, 1841 (1987); P.~D.~Drummond
and S.~J.~Carter, J.~Opt.~Soc.~Am.~B \textbf{4}, 1565 (1987); P.~D.~Drummond, R.~M.~Shelby, S.~R.~Friberg, and
Y.~Yamamoto, Nature \textbf{365}, 307 (1993). 
\bibitem{qsol2}S.~J.~Carter and P.~D.~Drummond, Phys.~Rev.~Lett \textbf{67}, 3757 (1991); M.~J.~Werner, Phys.~Rev.~A \textbf{54},
R2567 (1996). 
\bibitem{evapcool}P.~D.~Drummond and J.~F.~Corney, Phys.~Rev.~A \textbf{60}, R2661 (1999). 
\bibitem{becpf}M.~J.~Steel, M.~K.~Olsen, L.~I.~Plimak, P.~D.~Drummond, S.~M.~Tan, M.~J.~Collett, D.~F.~Walls, and R.~Graham,
Phys.~Rev.~A \textbf{58}, 4824 (1998). \end{thebibliography}
\end{document}